# Nonlinear charge transport in redox molecular junctions: a Marcus perspective.

Agostino Migliore[*,†] and Abraham Nitzan[*,†]

[†] *School of Chemistry, Tel Aviv University, Tel Aviv 69978, Israel. Phone: +972-3-6407634. Fax: +972-3-6409293.*

[*] E-mails: migliore@post.tau.ac.il and nitzan@post.tau.ac.il

**CORRESPONDING AUTHOR:** Agostino Migliore. School of Chemistry, Tel Aviv University, Tel Aviv 69978, Israel. Phone: +972-3-6407634. Fax: +972-3-6409293. E-mail: migliore@post.tau.ac.il

## Abstract

Redox molecular junctions are molecular conduction junctions that involve more than one oxidation state of the molecular bridge. This property is derived from the ability of the molecule to transiently localize transmitting electrons, implying relatively weak molecule-leads coupling and, in many cases, the validity of the Marcus theory of electron transfer. Here we study the implications of this property on the non-linear transport properties of such junctions. We obtain an analytical solution of the integral equations that describe molecular conduction in the Marcus kinetic regime and use it in different physical limits to predict some important features of nonlinear transport in metal-molecule-metal junctions. In particular, conduction, rectification and negative differential resistance can be obtained in different regimes of interplay between two different conduction channels associated with different localization properties of the excess molecular charge, without specific assumptions about the electronic structure of the molecular bridge. The predicted behaviors show temperature dependences typically observed in the experiment. The validity of the proposed model and ways to test its predictions and implement the implied control strategies are discussed.

## 1. Introduction.

Electron transport through molecular-scale systems and, in particular, across molecular junctions has been the subject of intensive experimental[1–4] and theoretical[1, 3, 5–7] investigations in the past few decades. A key challenge in this area is the detailed understanding of charge transport through single molecules[8] that are possibly embedded in a suitable environment.[1–4, 6, 9–11] In particular, the engineering of organic molecules and biomolecules, and tailoring of their properties by synthetic methods yields much more design flexibility than that permitted by typical inorganic materials. Extensive modeling studies are needed to guide such inquiries.[1]

In recent years, redox molecular junctions, that is junctions whose operation involves reversible transitions[12] between two or more oxidation states of the molecular bridge, have been the focus of many experimental[2, 11, 13–18] and theoretical[5, 7, 9, 19–21] studies, motivated by important features of nonlinear charge transport in such junctions and the control mechanisms offered by the correlation between their charging state and conductive properties.[2, 13] The ability of a molecular junction to switch between redox states is synonymous with the ability of the bridging molecule to localize an electron during the transmission process, which in turn depends on the relative alignment of the electrode Fermi levels and molecular energy levels (usually, the highest occupied molecular orbital (HOMO) and the lowest unoccupied molecular orbital (LUMO)) and on the interaction between the molecule and its thermal environment. The former depends on the bias potential and can be tuned by a gate potential, making it possible to achieve control of the molecular conductance.[2, 14]

While redox molecular conduction junctions can be envisioned in the gas phase or in vacuum, most studies of systems have focused on electrolyte solutions, where the solvent plays a central role both in assisting electron localization and in providing a convenient environment for electrochemical gating.[5, 14–18, 22–24] This has led to the locution "wet electronics"[14] and to the analysis of the transport properties of molecular junctions in terms of Marcus-type electron transfer (ET) [25–27] at the source-bridge and bridge-drain interfaces.

The "redox property" of a molecular conduction junction stems from two factors that must appear simultaneously. First and foremost is the existence of sequential ET processes that switch the molecule between two (or more) charging states. This requires that (a) relevant molecular orbitals, e.g. the LUMO and/or HOMO, are or can become localized about a molecular redox group (see Figure 1), and (b) electron localization is stabilized by suitable rearrangement and polarization of the nuclear environment. Dominating effect of such thermalization and localization processes mark the transition from tunneling to hopping, as was discussed extensively in the context of molecular electronic transport.[5, 7, 14, 16, 17]

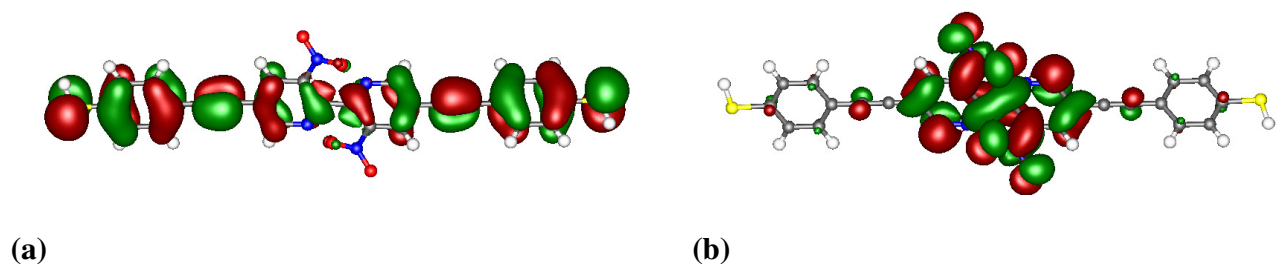

(a) (b)

**Figure 1.** Kohn-Sham **(a)** HOMO and **(b)** LUMO of the neutral BPDN.[28] The B3LYP hybrid functional and the 6-31g* basis set were used.

By themselves, these properties do not imply the redox character of a junction. A second condition is required: the existence of another transport channel, whose transmission efficiency is strongly affected by the change in the molecular redox state. Such a mechanism characterizes recent measurements of charge transport in quantum point contacts[29] and was also proposed[30] as a mechanism for negative differential resistance (NDR) in spin-blockaded transport through weakly coupled-double quantum dots.

It is important to emphasize the distinction implied by the second condition formulated above. Obviously, the first condition is sufficient for stating that conduction proceeds by consecutive transitions of the molecule between (at least) two redox states. However this condition by itself is not enough for observing different conduction properties associated with different oxidation states, as would be the case when the second condition is satisfied.

Thus, according to this picture, a "redox junction" involves at least two transmission channels: one whose relatively high efficiency is derived from a delocalized orbital that is well coupled to both leads,[31] and another, associated with a localized orbital that is weakly connected to the leads, for which (a) transient electron trapping can reinforce itself by subsequent nuclear relaxation, and (b) electron trapping, i.e. change in molecular redox state, strongly affects transport through the first channel. Many molecular bridges are indeed characterized by orbitals of both types that can potentially yield such behavior. Figure 1 displays, as an example, the HOMO and LUMO of bipyridyl-dinitro oligophenylene-ethynelene dithiol (BPDN) that show, even before nuclear relaxation is accounted for, considerable difference in localization properties.

The extension of the basic idea in ref 30 to weakly coupled (redox) molecular junctions can be accomplished by its appropriate combination with a theoretical framework for suitable description of the interfacial molecule-metal ET processes. This is the subject of the present work, where we provide a theoretical treatment that brings together the basics of the model from ref 30 and the Marcus theory for heterogeneous ET.[25–27] In section 2, we present a theoretical framework based on Marcus' ET theory for the study of charge transport in molecular junctions, valid when the molecular bridge is weakly coupled to the metal electrodes.[32] In sections 3-4 we show how the resulting theoretical model can be used to study the current-voltage response of redox molecular junctions, and to predict and study various NDR phenomena[33–39] and their dependence on properties of the junction. Section 5 concludes.

## 2. Theoretical Models.

**2.1. A single channel system.** Consider a molecular system coupled weakly to two metal electrodes $L$ and $R$ that are modeled as free electron reservoirs characterized by the respective chemical potentials. We start with the standard picture of a molecular bridge that comprises a single channel ("channel 1" in the following[40]), two-state system: an oxidized state $|A\rangle$ and a reduced state $|B\rangle$, with $N-1$ and $N$ electrons, respectively. In the unbiased junction, the corresponding energies, $E_A$ and $E_B$, and their difference, $\varepsilon_1 \equiv E_B - E_A$, depend on the proximity to the metal surfaces through the molecule-lead coupling, including image interactions.

When the molecule-electrode couplings are weak so that the timescale for electron transfer is long relative to that of thermal relaxation, electron transport through the molecular junction takes place through successive hopping processes, where each hopping step is associated with a rate obtained within the framework of the Marcus ET theory. The current-voltage characteristics of such a junction were extensively discussed in works by Kuznetsov and Ulstrup,[5, 7, 23] and is briefly reviewed here. The physics of ET between a metal electrode and a molecule differ from that between two molecular centers in two ways. First, solvent reorganization is affected by redox transitions on the molecule only: charging of the metal does not affect the solvent since this charge quickly delocalizes. Second, the energy balance includes the energy of the electron transferring to/from the Fermi sea, and depends on the molecular energy level alignment with respect to the lead Fermi energy and on the potential bias in the junction.

For definiteness we assign zero electrostatic potential to the molecular location, and denote the left (*L*) and right (*R*) electrode potentials by $\phi_L$ and $\phi_R$, respectively, so that a positive bias voltage corresponds to $\phi_R > \phi_L$. Assuming that the potential drops monotonically across the junction, we take, for positive bias voltage, $\phi_R > 0$ and $\phi_L < 0$, hence $\Delta\phi = \phi_R - \phi_L = \phi_R + |\phi_L|$. In the absence of nuclear relaxation, the molecule-electrode ET rates are given by[41]

$$R_{AB}^K = \int_{-\infty}^{\infty} dE\, \gamma_1^K(E) f_K(E) \delta(E - \varepsilon_1) = \gamma_1^K(\varepsilon_1) f_K(\varepsilon_1) \tag{1a}$$

and

$$R_{BA}^K = \int_{-\infty}^{\infty} dE\, \gamma_1^K(E)[1 - f_K(E)]\delta(\varepsilon_1 - E) = \gamma_1^K(\varepsilon_1)[1 - f_K(\varepsilon_1)], \tag{1b}$$

where *K* = *L* or *R*. *AB* and *BA* refer to the processes $A \rightarrow B$ (electron injection into the molecule) and $B \rightarrow A$ (electron removal from the molecule), respectively. In eqs 1a-b, it is

$$f_K(E) = \frac{1}{\exp\left(\dfrac{E - \mu + e\phi_K}{k_B T}\right) + 1}, \tag{2}$$

is the Fermi function describing the electronic occupation in electrode *K* ($\mu$, $e$, $k_B$, and *T* are chemical potential in the absence of bias, magnitude of the electron charge, Boltzmann constant and temperature, respectively), and $\gamma_1^K$ is given in terms of the metal-molecule coupling $V_{AB}^K$ and the density of single electron states in the metal, $\rho_K(E)$, by the golden rule formula

$$\gamma_1^K(E) = \frac{2\pi}{\hbar}|V_{AB}^K|^2 \rho_K(E). \tag{3}$$

Taking nuclear relaxation into account, eqs 1a-b are replaced by

$$R_{AB}^K = \int_{-\infty}^{\infty} dE\, \gamma_1^K(E) f_K(E) F_1(E - \varepsilon_1) \tag{4a}$$

and

$$R_{BA}^K = \int_{-\infty}^{\infty} dE\, \gamma_1^K(E)[1 - f_K(E)] F_1(\varepsilon_1 - E), \tag{4b}$$

where, in the semiclassical limit, the function $F_1$ has the form[25]

$$F_1(u) = \frac{1}{2\sqrt{\pi \lambda_1 k_B T}} \exp\left[-\frac{(u - \lambda_1)^2}{4\lambda_1 k_B T}\right]. \tag{5}$$

$\lambda_1$ is the value of the reorganization energy for the ET steps involved in the given transport channel, and is determined by the bridge charging state associated with this channel. It is the free energy released by relaxation of the nuclear environment to its stable configuration following a sudden transition between the charge distributions associated with the $A$ and $B$ molecular states. In eq 1 and eq 4, the energy integration should be over the metallic band, and is extended to $\pm\infty$ under the assumption that the integrand is well included in this band.

Disregarding the energy dependence of $\gamma_1^K(E)$ and changing integration variable, eqs 4a-b become[42]

$$R_{AB}^K = \frac{\gamma_1^K}{2\sqrt{\pi \lambda_1 k_B T}} \int_{-\infty}^{\infty} dE\, \frac{1}{\exp\left(\frac{E}{k_B T}\right) + 1} \exp\left[-\frac{(E + \mu - \varepsilon_1 - e\phi_K - \lambda_1)^2}{4\lambda_1 k_B T}\right] \tag{6a}$$

and

$$R_{BA}^K = \frac{\gamma_1^K}{2\sqrt{\pi \lambda_1 k_B T}} \int_{-\infty}^{\infty} dE\, \frac{1}{\exp\left(-\frac{E}{k_B T}\right) + 1} \exp\left[-\frac{(-E - \mu + \varepsilon_1 + e\phi_K - \lambda_1)^2}{4\lambda_1 k_B T}\right]. \tag{6b}$$

Expressions (6) have been extensively used in theoretical analyses of electrochemical processes,[43] including electron transport in electrochemical molecular junctions.[20, 21, 44]

In this work, in departure from standard treatments we consider the general expressions obtained from evaluating these integrals (see Appendix A):

$$R_{AB}^{K} = \frac{\gamma_1^K}{4} S(\lambda_1, T, \alpha_K) \exp\left[-\frac{(\alpha_K - \lambda_1)^2}{4\lambda_1 k_B T}\right], \qquad R_{BA}^{K} = \frac{\gamma_1^K}{4} S(\lambda_1, T, \alpha_K) \exp\left[-\frac{(\alpha_K + \lambda_1)^2}{4\lambda_1 k_B T}\right] \qquad (7a)$$

where

$$\alpha_K \equiv \mu - \varepsilon_1 - e\phi_K, \qquad (7b)$$

$$S(\lambda, T, \alpha) = \sum_{n=0}^{N} \frac{1}{2^n} \sum_{j=0}^{n} (-1)^j \binom{n}{j} \left[\chi_j(\lambda, T, \alpha) + \chi_j(\lambda, T, -\alpha)\right], \qquad (7c)$$

$$\chi_j(\lambda, T, \alpha) = \exp\left\{\frac{[(2j+1)\lambda + \alpha]^2}{4\lambda k_B T}\right\} \mathrm{erfc}\left[\frac{(2j+1)\lambda + \alpha}{2\sqrt{\lambda k_B T}}\right], \qquad (7d)$$

and the limit superior $N$ truncates the otherwise infinite sums.

In eqs 7a-d, the ET rates are expressed as summations over analytic (or entire in the complex plane) functions that depend on the values of the physical parameters $\lambda$, $\varepsilon_1$, $\mu$ and $T$. These expressions open the way to systematic approximations of the integrals in eq 6. For example, Hale's approximation to such integrals,[45] used in previous studies of electrochemical redox reactions at solution-metal interfaces, amounts to retain only $\chi_0(\lambda_1, T, -\alpha_L)$ and $\chi_0(\lambda_1, T, \alpha_R)$ in $R_{AB}^L$ and $R_{BA}^R$, respectively. As detailed in the Supporting Information, the theoretical analysis of Appendix A establishes such approximation under physical constraints weaker than those used in ref 45, and sets the general limits of its applicability in terms of the reorganization energy and the applied bias. Another approximation adopted by Marcus[26] considers only the ET processes to and from the electrode Fermi levels. This leads, essentially, to the common Gaussian (as a function of the applied voltage) factor of eqs 7a, and can appropriately describe the molecule-metal ET when the overpotential[43] and/or voltage[20] are significantly smaller than the reorganization energy. The same kind of rate expression was used for bridge-metal ET in redox molecular junctions, leading to current-voltage characteristics where the current decreases at sufficiently high bias voltages,[5, 7, 46] while its behavior at small biases depends on the system structure and coupling parameters. Still another approximation amounts to disregarding the term quadratic in the bias-dependent reaction free energy in the argument of the Marcus-type free energy factor of eqs 6. This approximation is possible, over a suitably small bias range, if the reorganization energy is much larger than the reaction free energy for the given ET reaction and leads to rates with exponential dependence on both the reorganization and reaction free energies.[20, 21, 41, 44] All these approximations are, indeed, special cases of eqs 7a, which can be exploited to obtain further useful approximations (see, e.g., Figure S1 in the Supporting Information). We shall see that eqs 7a-d provide convenient ET rate expressions also for describing the $I$-$\Delta\phi$ characteristics of electrochemical molecular conduction junctions.

For the analysis below, it will be useful to define the electron-exchange threshold voltage (EETV) at each given contact as the electrode-molecule potential difference at which the pertinent transition rate reaches half of its limiting (high-voltage) value. In a positively biased junction ($\phi_L < 0 < \phi_R$), electrons move in the direction $L \rightarrow M \rightarrow R$. The corresponding EETV values are approximately given by (see Appendix A)

$$-e\phi_L^{EETV(AB)} = \lambda_1 + \varepsilon_1 - \mu \tag{8a}$$

and

$$e\phi_R^{EETV(BA)} = \lambda_1 - (\varepsilon_1 - \mu). \tag{8b}$$

These thresholds depend on the reorganization energy and both need to be overcome to obtain efficient transport through the junction. For example, for $\varepsilon_1 - \mu > 0$, eqs 8a-b imply that $-\phi_L^{EETV(AB)} > \phi_R^{EETV(BA)}$. So, for symmetrical potential distribution, $-\phi_L = \phi_R = \Delta\phi/2$, the transition $L \rightarrow M$ is rate limiting. In these situations, the effect of the reorganization energy can become non-trivial. For example, if $\lambda_1 < \varepsilon_1 - \mu$, the $M \rightarrow R$ transition is enabled for any positive voltage while $\lambda_1 > \varepsilon_1 - \mu$ implies that there is a threshold potential. In the latter case, for an asymmetric potential distribution such that $-\phi_L \gg \phi_R$, the $M \rightarrow R$ process may become rate limiting.

It should also be noted also that when the EETVs are surpassed for the forward ($L \rightarrow M \rightarrow R$) transfer, the opposite rates are negligible. This follow from (see Appendix A)

$$R_{BA}^{K} = R_{AB}^{K} \exp\left(-\frac{\alpha_K}{k_B T}\right), \tag{9}$$

For example, for $\lambda_1$ of the order of 1 eV, eqs 7-9 show that the backward rates $R_{BA}^{L}$ and $R_{AB}^{R}$ are negligible at voltages such that the forward rates $R_{AB}^L$ and $R_{BA}^R$, hence the current through the given channel, are not yet appreciable. We will see that the backward rates can play a role when more than one channel is involved in the junction transport process.

Given the rates (7), the steady-state current can be obtained from the classical rate equations for the probabilities $P_A$ and $P_B = 1 - P_A$ for the molecule to be in states $|A\rangle$ and $|B\rangle$, respectively:

$$\frac{dP_A}{dt} = -P_A R_{AB} + P_B R_{BA} = 0, \tag{10}$$

where $R_{AB} = R_{AB}^L + R_{AB}^R$ and $R_{BA} = R_{BA}^L + R_{BA}^R$. Using eq 9, this leads to the following equation for the steady-state current $I$:

$$\frac{I}{e} = \frac{R_{AB}^L R_{BA}^R - R_{AB}^R R_{BA}^L}{R_{AB} + R_{BA}} = \frac{R_{AB}^L R_{BA}^R \left[1 - \exp\left(-\frac{e\Delta\phi}{k_B T}\right)\right]}{R_{AB}^L \left[1 + \exp\left(-\frac{\mu - \varepsilon_1 - e\phi_L}{k_B T}\right)\right] + R_{BA}^R \left[1 + \exp\left(-\frac{\varepsilon_1 - \mu + e\phi_R}{k_B T}\right)\right]}. \tag{11a}$$

For positive bias ($\phi_L < 0$, $\phi_R > 0$), the current becomes significant at bias voltages that satisfy both eqs 8. When the backward ET rates can be disregarded, this yields

$$\frac{I}{e} \cong \frac{R_{AB}^L R_{BA}^R}{R_{AB}^L + R_{BA}^R} \left[1 - \exp\left(-\frac{e\Delta\phi}{k_B T}\right)\right]. \tag{11b}$$

Moreover, the exponential in the brackets is negligible when the bias is such that $e\Delta\phi \gg k_B T$, leading to

$$\frac{I}{e} \cong \frac{R_{AB}^L R_{BA}^R}{R_{AB}^L + R_{BA}^R}, \tag{12}$$

Figure 2 illustrates some results based on the full steady state eq 11a and the approximation of eq 12. As shown in Figure 2a, electronic-nuclear coupling (expressed by the reorganization energy) affects both the EETVs (thus, the threshold voltage across the junction for the onset of appreciable current) and the voltage width of the threshold region. It is seen that the larger is the reorganization energy, the smaller is the effect of a given potential change on the energetics of the system and the larger is the width of the current rise along the bias sweep (see Figure 2a). This is, indeed, rigorously quantified by eqs 7 (see also the approximate eqs A5-6). Figure 2 also shows that essentially the behavior of the current, including its rise to the high-bias plateau value, is well described by eq 12. This extends significantly the applicability of eq 12 compared to its use in ref 30. Obviously, the exact expression of eq 11a is to be used in order to obtain the exact current-voltage response at low bias (visible in the semilog plots of the insets), in particular for evaluating the initial slope $dI/d(\Delta\phi)|_{\Delta\phi=0}$, and at high temperature.

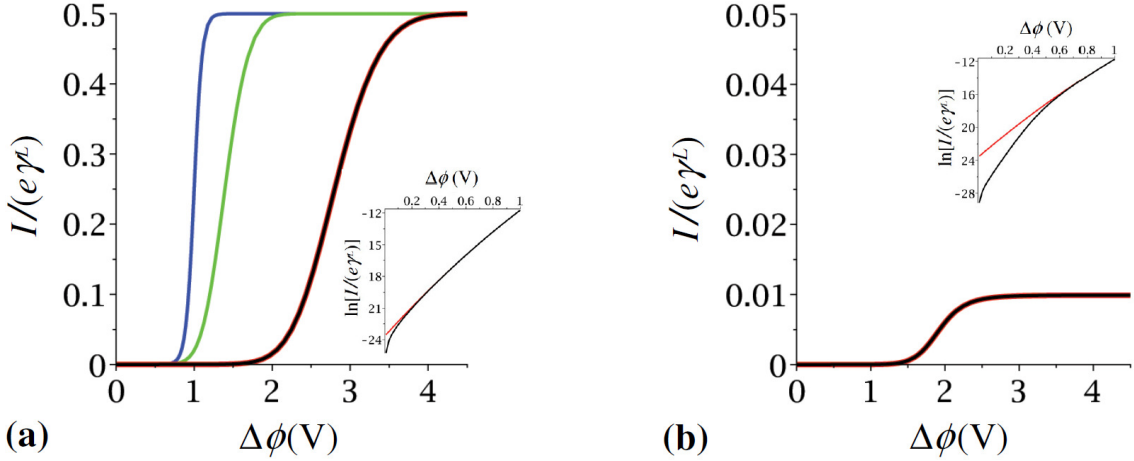

**Figure 2.** $I/(e\gamma^L)$ plotted against $\Delta\phi$, from eq 11a (blue, green and black) or eq 12 (red), for a one-channel system characterized by the parameters: $T = 298\,\text{K}$, $-\phi_L = \phi_R = \Delta\phi/2$, $\varepsilon_1 - \mu = 0.5\,\text{eV}$, and **(a)** $\gamma_1^L = \gamma_1^R$ (symmetrical contacts) and $\lambda_1 = 0\,\text{eV}$ (blue line), 0.25 eV (green), 1 eV (black and red); **(b)** $\gamma_1^R = 0.01\gamma_1^L$ (markedly asymmetric contacts) and $\lambda_1 = 1.0\,\text{eV}$.

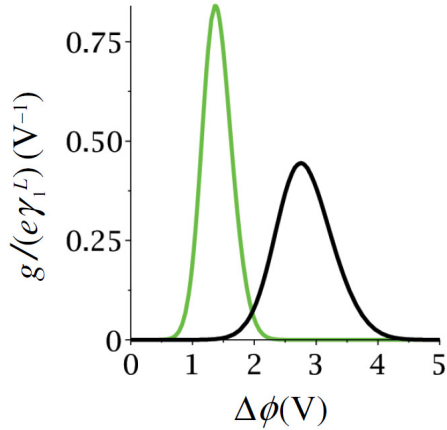

**Figure 3.** $g/(e\gamma_1^L)$ versus $\Delta\phi$, computed by the expression of the current in eq 11a, for the same model parameters as in Figure 2a. The same color code is employed.

The differential conductance $g = dI/d(\Delta\phi)$ (Figure 3) shows, as a function of the bias, a peak[41] which is approximately located at the voltage $\Delta\phi = 2|\phi_L| = 2(\lambda_1 + \varepsilon_1 - \mu)/e$. Similar peaks in the $I$-$\Delta\phi$ curve are an experimental signature of weak couplings between bridge and leads (see, e.g., refs 47 and 48).

**2.2. A two channel system.** Next, consider a molecule characterized by three states, $|A\rangle$, $|B\rangle$ and $|C\rangle$, which are accessible within the bias range of interest. In particular, following ref 30, we focus on the case where $|A\rangle$ and $|B\rangle$ correspond, as above, to two different molecular charging states while state $|C\rangle$ has energy $E_C = E_A + \varepsilon_2$ and the same number of electrons as $|B\rangle$. For example, $|B\rangle$ and $|C\rangle$ may be the ground and first-excited states, respectively, of the molecular system with one excess electron, while states corresponding to a doubly charged molecule are assumed to be too high in energy to contribute in the voltage range considered here. The $A \leftrightarrow C$ transition thus constitutes another transmission channel, channel 2, characterized by a reorganization energy $\lambda_2$. The analogs of eqs 8a-b for this channel are then

$$-e\phi_L^{EETV(AC)} = \lambda_2 + \varepsilon_2 - \mu, \tag{13a}$$

$$e\phi_R^{EETV(CA)} = \lambda_2 - (\varepsilon_2 - \mu). \tag{13b}$$

The system is now characterized by two transport channels that involve the $A \leftrightarrow B$ and $A \leftrightarrow C$ ET processes. Kinetic equations analogous to (10) yield (see Appendix B)

$$I = \frac{1}{1 + \dfrac{R_{AB}^L + R_{BA}^R \exp\left(-\dfrac{\varepsilon_1 - \mu + e\phi_R}{k_B T}\right)}{R_{BA}^R + R_{AB}^L \exp\left(-\dfrac{\mu - \varepsilon_1 - e\phi_L}{k_B T}\right)} + \dfrac{R_{AC}^L + R_{CA}^R \exp\left(-\dfrac{\varepsilon_2 - \mu + e\phi_R}{k_B T}\right)}{R_{CA}^R + R_{AC}^L \exp\left(-\dfrac{\mu - \varepsilon_2 - e\phi_L}{k_B T}\right)}} \\ \times \left\{ I_{AB} \left[ 1 + \dfrac{R_{AB}^L + R_{BA}^R \exp\left(-\dfrac{\varepsilon_1 - \mu + e\phi_R}{k_B T}\right)}{R_{BA}^R + R_{AB}^L \exp\left(-\dfrac{\mu - \varepsilon_1 - e\phi_L}{k_B T}\right)} \right] + I_{AC} \left[ 1 + \dfrac{R_{AC}^L + R_{CA}^R \exp\left(-\dfrac{\varepsilon_2 - \mu + e\phi_R}{k_B T}\right)}{R_{CA}^R + R_{AC}^L \exp\left(-\dfrac{\mu - \varepsilon_2 - e\phi_L}{k_B T}\right)} \right] \right\}. \tag{14}$$

where $R_{AC}^L$ and $R_{CA}^R$ are the rates of the forward (left to right) ET processes that involve $|C\rangle$, given by expressions similar to the first eq 7a for $K = L$ and the second eq 7a for $K = R$, respectively, except that $B$, $\varepsilon_1$, $\lambda_1$, $\gamma_1^L$ and $\gamma_1^R$ are replaced by $C$, $\varepsilon_2$, $\lambda_2$, $\gamma_2^L$ and $\gamma_2^R$.[49] $I_{AB}$ and $I_{AC}$ are the currents that would be obtained if $|C\rangle$ or $|B\rangle$, respectively, were inaccessible. Note however that when both channels are active $I_{AB}$ and $I_{AC}$ do not simply combine additively to give the overall current, because each of the bias-dependent probabilities depends on the existence of three molecular electronic states.

When the backward ET rates can be disregarded, eq 14 is simplified to give the analog of eq 12:

$$\frac{I}{e} \cong \frac{R^L_{AB} + R^L_{AC}}{1 + \frac{R^L_{AB}}{R^R_{BA}} + \frac{R^L_{AC}}{R^R_{CA}}}. \tag{15}$$

As before, we shall see that eq 15 has a broad range of validity and can describe the full current-voltage response in several cases (e.g., see Figure 4), not only in the plateau regime as asserted in ref 30.

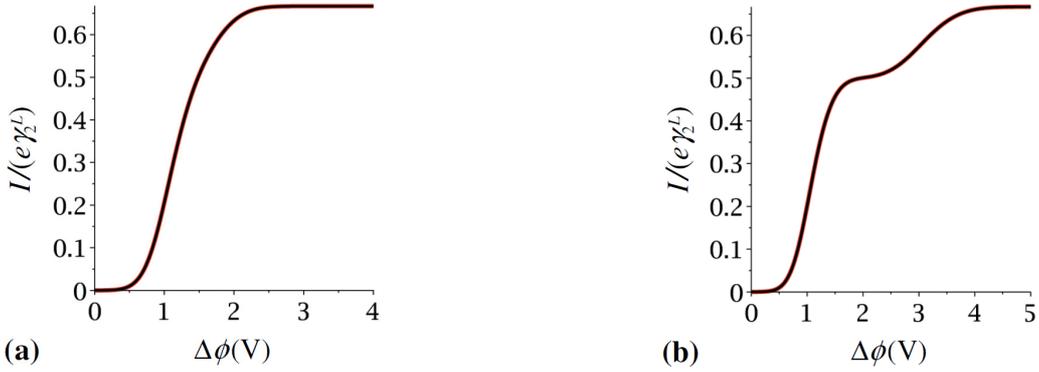

**Figure. 4.** $I/(e\gamma^L_2)$ versus $\Delta\phi$, computed from eq 14 (black) and eq 15 (red) with model parameters $T = 298\,\mathrm{K}$, $-\phi_L = \phi_R = \Delta\phi/2$, $\varepsilon_1 - \mu = 0.2\,\mathrm{eV}$, $\varepsilon_2 - \mu = 0.4\,\mathrm{eV}$, $\gamma^L_1 = \gamma^R_1 = \gamma^L_2 = \gamma^R_2$, $\lambda_1 = 0.4\,\mathrm{eV}$, and **(a)** $\lambda_2 = 0.5\,\mathrm{eV}$ or **(b)** $\lambda_2 = 1.2\,\mathrm{eV}$.

If channels 1 and 2 are accessed sequentially, then the current is described by eqs 11 when channel 2 is not involved. Above the threshold bias for channel 2, the two channels compete in determining *I*. By comparing eqs 8 and 13, and assuming for definiteness that $\varepsilon_2 - \mu > 0$, eqs A5-6 lead to the following condition for sequential access when $|\phi_L| = \phi_R = \Delta\phi/2$:

$$\left|\phi_L^{EETV(AC)}\right| - \max\left(\left|\phi_L^{EETV(AB)}\right|, \phi_R^{EETV(BA)}\right) \geq \frac{2s}{e}\sqrt{k_B T}\left(\sqrt{\lambda_1} + \sqrt{\lambda_2}\right), \tag{16}$$

where *s* is a positive real number such that $\mathrm{erfc}(u)$ can be considered negligible for $u \geq s$, namely, the EETV for appreciable occupation of the *C* state is larger than the maximum EETV involved in the electron transport through channel 1 by at least the sum of the widths of the voltage ranges where the respective transition rates rise up. The right-hand side of eq 16 is a consequence of the smoothed voltage dependence of the current in the presence of nuclear relaxation effects.

Figure 4 shows two examples of *I*-$\Delta\phi$ characteristics resulting from the above equations. In Figure 4a, the current steps associated with the two channels are merged into a single step, while two distinct steps are seen in Figure 4b because of the different reorganization energies associated with the two channels.

We conclude this section with the following remarks:

(i) Despite the excellent performance of eq 15 in several cases, as exemplified in Figure 4, there are also cases where it fails in predicting the current-voltage characteristics in significant bias ranges and regarding important features (see section 4) so that eq 14 must be used.

(ii) In the classical rate picture (diagonal density matrix) assumed here, channels 1 and 2 provide distinct transmission routes: An electron transmission event either involves one or the other. Moreover, they are mutually exclusive: the outer molecular orbital pertaining to state $|B\rangle$ cannot be occupied if the one pertaining to state $|C\rangle$ is already occupied, and vice versa. If the molecule is trapped in state $|C\rangle$, transmission via channel 1 cannot take place. Physically, this implies as was already stated above, that in the voltage range of interest the molecule cannot enter a doubly charged state that accommodates the two electrons in orbitals 1 and 2.

(iii) With reference to eq 16, it is worth noting that for $\lambda_2 \neq 0$, if the threshold voltage for efficient *C* → *A* transition with ET to the *R* lead is larger than that for the reverse process involving the *L* contact, then channel 2 is enabled at higher biases than those required for the access of state $|C\rangle$. Therefore, in a given voltage range, the transferring electron can be trapped in $|C\rangle$ rather than being carried across the junction through the *AB* channel. As a result, the *AC* transport channel can be "accessed" and thus affect the current while it is still not able to convey appreciable current. Hence, the possible mismatch between the threshold biases for the access of state $|C\rangle$ and the activation of the corresponding transport channel needs generally to be considered along with eq 16 in order to understand phenomena such as NDR in terms of the interplay between the dynamics of the two channels (see section 4).

(iv) In the notation, an ideal gate voltage (that affects only energetics in the bridging molecule), $V_g$, is entered by the substitution $\phi_K \to \phi_K + V_g$, or the equivalent substitution $\{\varepsilon_1, \varepsilon_2\} \to \{\varepsilon_1 + eV_g, \varepsilon_2 + eV_g\}$, in eqs 7 and the corresponding equations for channel 2, for both $K = L, R$.

## 3. Current rectification and control in the one-channel model.

It has long been known[50] that asymmetric potential distribution in a junction (in our model, $-\phi_L \neq \phi_R$) leads to rectification behavior. While the principle remains, its manifestation depends on the nuclear reorganization. To see this, let us consider the two-state model discussed in section 2.1, and, following ref 50, characterize the voltage distribution in the junction by

$$\phi_L = -w\Delta\phi, \quad \phi_R = (1-w)\Delta\phi, \tag{17}$$

where $0 \leq w \leq 1$ is the voltage division factor.[51] By making explicit the dependence of the current on $w$ and $\Delta\phi$, eq 11a takes the form

$$\frac{I(\Delta\phi; w)}{e} = -\frac{I(-\Delta\phi; 1-w)}{e} = \frac{R_{AB}^L(w\Delta\phi) R_{BA}^R((1-w)\Delta\phi)\left[1 - \exp\left(-\frac{e\Delta\phi}{k_B T}\right)\right]}{R_{AB}^L(w\Delta\phi)\left[1 + \exp\left(-\frac{\mu - \varepsilon_1 + w\Delta\phi}{k_B T}\right)\right] + R_{BA}^R((1-w)\Delta\phi)\left\{1 + \exp\left[-\frac{\varepsilon_1 - \mu + (1-w)\Delta\phi}{k_B T}\right]\right\}}. \tag{18}$$

Consistently with the result of ref 50, eq 18 implies that, for a given junction with a fixed $w$, $|I(\Delta\phi; w)/I(-\Delta\phi; w)| \neq 1$ unless $w = \frac{1}{2}$. However, the molecular and solvent reorganization following each ET process affects this behavior in a significant way, as discussed below.

For $w \neq \frac{1}{2}$, insertion of eqs 17 into eqs 8a-b, as well as into the corresponding conditions for negative bias, namely

$$e\phi_L^{EETV(BA)} = \lambda_1 - (\varepsilon_1 - \mu), \tag{19a}$$

$$-e\phi_R^{EETV(AB)} = \lambda_1 + \varepsilon_1 - \mu, \tag{19b}$$

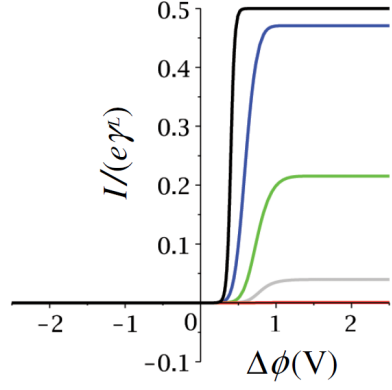

**Figure. 5.** $I/(e\gamma^L)$ versus $\Delta\phi$ obtained from eq 18 with parameters $T = 298\,\mathrm{K}$, $\gamma_1^L = \gamma_1^R$, $w = 1-10^{-5}$, $\varepsilon_1 - \mu = 0.4\,\mathrm{eV}$ and $\lambda_1 = 0\,\mathrm{eV}$ (black), 0.25 eV (blue), 0.5 eV (green), 0.75 eV (grey), 1.25 eV (red).

leads to the threshold voltages for enabling the molecular channel under positive and negative voltages:

$$\Delta\phi > \max\left[-\frac{\phi_L^{EETV(AB)}}{w}, \frac{\phi_R^{EETV(BA)}}{1-w}\right] = \max\left[\frac{1}{w}\frac{\lambda_1 + \varepsilon_1 - \mu}{e}, \frac{1}{1-w}\frac{\lambda_1 - (\varepsilon_1 - \mu)}{e}\right] \quad (\Delta\phi > 0), \quad (20a)$$

$$-\Delta\phi > \max\left[\frac{e\phi_L^{EETV(BA)}}{w}, -\frac{e\phi_R^{EETV(AB)}}{1-w}\right] = \max\left[\frac{1}{w}\frac{\lambda_1 - (\varepsilon_1 - \mu)}{e}, \frac{1}{1-w}\frac{\lambda_1 + \varepsilon_1 - \mu}{e}\right] \quad (\Delta\phi < 0), \quad (20b)$$

For $\varepsilon_1 - \mu > 0$ and $\lambda_1 = 0$, these conditions read

$$\Delta\phi > \frac{1}{w}\frac{\varepsilon_1 - \mu}{e} \quad (\Delta\phi > 0), \quad (21a)$$

$$-\Delta\phi > \frac{1}{1-w}\frac{\varepsilon_1 - \mu}{e} \quad (\Delta\phi < 0), \quad (21b)$$

which is identical to eq 2 of ref 50. Eq 20 represents the generalization of this condition to cases where nuclear reorganization plays a role in the charge transfer processes. As shown in Figure 5, for $w$ close to unity, the rectification effect, quantified for any given $\Delta\phi$ by $|I(\Delta\phi > 0; w)/I(\Delta\phi < 0; w)|$, depends on the value of $\lambda_1$. The difference between the limiting situations, among those illustrated, is easily understood from a comparison between eqs 20 and 21. For $\lambda_1 = 0$ the threshold imposed by eq 21b is not overcome in the explored negative bias range and no appreciable current is obtained for $\Delta\phi < 0$. Instead, the current reaches its high-voltage maximum value under positive bias, in agreement with eq 21a (see black characteristic). On the contrary, for $\lambda_1$ sufficiently larger than $\varepsilon_1 - \mu$ (as, e.g., in the case of the red characteristics), neither of conditions in eq 20 is achieved in the spanned voltage range, and the current remains negligible also under positive bias.

In order to gain a better understanding of the modulation of threshold bias and plateau current by the reorganization energy, as shown in Figure 5, we consider eq 18 in the limiting cases $w \to 0$ and $w \to 1$:

$$\lim_{w \to 0} \frac{I(\Delta\phi; w)}{e} \cong \frac{R_{AB}^L(0) R_{BA}^R(\Delta\phi)\left[1 - \exp\left(-\frac{e\Delta\phi}{k_B T}\right)\right]}{R_{AB}^L(0)\left[1 + \exp\left(\frac{\varepsilon_1 - \mu}{k_B T}\right)\right] + R_{BA}^R(\Delta\phi)\left[1 + \exp\left(-\frac{\varepsilon_1 - \mu + e\Delta\phi}{k_B T}\right)\right]} \quad (22a)$$

$$\lim_{w \to 1} \frac{I(\Delta\phi; w)}{e} \cong \frac{R_{AB}^L(\Delta\phi) R_{BA}^R(0)\left[1 - \exp\left(-\frac{e\Delta\phi}{k_B T}\right)\right]}{R_{AB}^L(\Delta\phi)\left[1 + \exp\left(-\frac{\mu - \varepsilon_1 + e\Delta\phi}{k_B T}\right)\right] + R_{BA}^R(0)\left[1 + \exp\left(\frac{\mu - \varepsilon_1}{k_B T}\right)\right]}. \quad (22b)$$

Eqs 7 or eqs A5-6 implies that the value of $R_{AB}^L$ at zero bias is negligible compared to its maximum, $\gamma^L$, whenever $\lambda_1 \gg k_B T$. This makes the first limit negligible. In contrast, $R_{AB}^L$ takes the plateau value, $\gamma^L$, sufficiently above the threshold bias given by eq 8a. On the other hand, $R_{BA}^R(0)$ is of the order of $\gamma^R$ for $|\lambda_1 - (\varepsilon_1 - \mu)| < 2\sqrt{\lambda_1 k_B T}$, as shown by eq 7 or eq A6. Thereby, for the common situation $\varepsilon_1 - \mu \gg k_B T$, the insertion of eqs 7 into eqs 22 leads to

$$\begin{cases} \lim_{w \to 0} \dfrac{I(\Delta\phi; w)}{e} \cong 0, & \Delta\phi \geq 0 \quad (23a) \\ \lim_{w \to 1} \dfrac{I(\Delta\phi; w)}{e} \cong \dfrac{\gamma^L \gamma_1^R S(\lambda_1, T, \mu - \varepsilon_1) \exp\left[-\dfrac{(\mu - \varepsilon_1 + \lambda_1)^2}{4\lambda_1 k_B T}\right]}{4\gamma^L + \gamma_1^R S(\lambda_1, T, \mu - \varepsilon_1) \exp\left[-\dfrac{(\mu - \varepsilon_1 + \lambda_1)^2}{4\lambda_1 k_B T}\right]}, & \Delta\phi - \dfrac{\lambda_1 + \varepsilon_1 - \mu}{e} \gg 2\sqrt{\lambda_1 k_B T}. \quad (23b) \end{cases}$$

Eq 23a and the first equality in eq 18 imply a negligible current under $\Delta\phi \leq 0$ for the cases with $w \to 1$ represented in Figure 5. On the contrary, eq 23b establishes the threshold voltage for the onset of appreciable current and its limiting value in any reasonable bias range.[52] Both quantities are tuned by the value of the reorganization energy and depend also on the temperature. In particular, notice that the plateau current is generally different from the value $e\gamma^L\gamma^R/(\gamma^L+\gamma^R)$ obtained when $\lambda_1=0$.[30]

We conclude by reiterating the important observation that, as with other conduction properties, also the rectification behavior of asymmetric redox junctions is affected by the dielectric relaxation of the solvent environment, which often provides the main contribution to the reorganization energy.

## 4. The three-state model: channel competition and NDR.

A richer array of behaviors results from the simultaneous operation of the two transport channels in the model described in Section 2.2. Indeed, Muralidharan and Datta[30] have shown that such junction can display NDR if the *AC* channel, which is accessed at higher bias voltage, has a blocking character. This can happen if the system can go relatively easily into state $|C\rangle$, but, once there, takes a long time to switch back to $|A\rangle$. Since this route competes with the conducting *AB* pathway, conductance goes down at the threshold bias voltage for populating $|C\rangle$. We note that stabilization by solvent reorganization following the occupation of state $|C\rangle$ is one way to impart a blocking character to this state.

In Appendix B we reproduce the kinetic analysis of ref 30, which leads to the condition

$$\frac{1}{\gamma_2^R} > \frac{1}{\gamma_1^L} + \frac{1}{\gamma_1^R} \tag{24}$$

for current collapse or NDR under increasing positive bias voltage, assuming sequential access of the two transport channels and disregarding the backward electron transfer in the bias range where NDR appears. In this section, we show that the nuclear reorganization energies that characterize the two conduction modes play an important role in the competitive transport through the *AB* and *AC* channels. Before discussing the effects of $\lambda_1$ and $\lambda_2$, it is useful to consider the general mechanism of channel competition by means of the following expression of the current:

$$I = \frac{1}{1+\dfrac{R_{AB}}{R_{BA}}+\dfrac{R_{AC}}{R_{CA}}}\left[I_{AB}\left(1+\frac{R_{AB}}{R_{BA}}\right)+I_{AC}\left(1+\frac{R_{AC}}{R_{CA}}\right)\right] = (P_A+P_B)I_{AB}+(P_A+P_C)I_{AC}, \tag{25}$$

obtained by combining eqs B6 and B7. The first term in the last expression is the probability that the molecular bridge is in state *A* or *B* (hence channel 1 is active) multiplied by $I_{AB}$, which would be the pertinent current in the absence of state *C*. The second term is similarly described for channel 2. $P_A$ is clearly involved in the occupation probabilities of both transport channels, while $P_B$ and $P_C$ correspond to mutually exclusive events and must satisfy the constraint $P_B + P_C = 1 - P_A$. Overall, eq 25 expresses the non-additivity of $I_{AB}$ and $I_{AC}$, due to the mutually exclusive nature of the transport through the corresponding channels. At biases such that $P_C$ and thus $I_{AC}$ are negligible, $I_{AB}$ equals the total current through the junction (e.g., see Figure 6b below). In contrast, in general, $I_{AC}$ does not approximate the actual current at any voltage. It is interesting to note that the presence of channel 2 can significantly affect the current even at biases where both the $A \rightarrow C$ and $C \rightarrow A$ transition rates are negligible. This occurs if $R_{CA} \ll R_{AC}$ or, at high enough bias, $R_{CA}^R \ll R_{AC}^L$, even if $R_{AC}^L$ is much smaller than $R_{AB}^L$ and $R_{BA}^R$. In this case eq B6 gives $P_A \cong 0$, $P_B \cong 0$, $P_C \cong 1$, and eq 25 yields $I \cong I_{AC} \cong 0$. This is easily understood by considering a gedanken experiment where the system is observed for a virtually infinite time at each applied voltage $\Delta \phi$. Many $A \leftrightarrow B$ and $A \leftrightarrow C$ transitions take place at that $\Delta \phi$, and the bias-dependent probabilities of $|A\rangle$, $|B\rangle$ and $|C\rangle$ can be measured. Then a high $R_{AC}^L / R_{CA}^R$ value implies that the system spends most time in state $|C\rangle$, so that $P_A$ and $P_B$ are negligible and the current is given by the very small $I_{AC}$.

*4.1. Effect of the channel reorganization energies on NDR phenomena.* Consider now the effect of nuclear relaxation on such NDR phenomena.. Within our present framework they arise from the dependence of the rates that appear in eq 25 on the reorganization energies $\lambda_1$ and $\lambda_2$. It should be noted that in many circumstances one may expect $|\lambda_2 - \lambda_1| < \lambda_1, \lambda_2$, because both reorganization energies correspond to a change in molecular charge between states $|A\rangle$ and either $|B\rangle$ or $|C\rangle$ while the difference $\lambda_2 - \lambda_1$ corresponds to transition between two states of the same charge. However, one can envisage several cases where $|\lambda_2 - \lambda_1|/\lambda_1$ is significantly larger than unity. For example, this is the case in systems such as dithiophene-tetrathiafulvalene[53] and alumina,[54] in which the reorganization energies are dominated by intramolecular relaxation. This may also happen when the *B* and *C* states involve very different distributions of the excess charge on the molecule: the reorganization energy is relatively small and not appreciably affected by the presence of a redox site if the transferring electron is delocalized over the molecule,[55] while it is expected to be much larger if the charge localizes at the redox center.[56] Furthermore, metal-molecule-metal electron transfers mediated by different redox moieties in the molecule (e.g., in azurin[57]) can also imply very different $\lambda_1$ and $\lambda_2$ values. In such cases we expect large effect of the reorganization energies on the resulting NDR effects as shown in Figures 6 and 7 below.

Figure 6 shows current-voltage characteristics for systems with the same molecular level and coupling parameters, but with different reorganization energies. Although the junction parameters satisfy eq 24 in all cases, NDR is seen only in panels a and b. The characteristics of Figures 6a-b for nonzero $\lambda_1$ and $\lambda_2$ (black lines) show that the current peak broadens and shifts to higher bias voltages compared to the case $\lambda_1 = \lambda_2 = 0$ (blue line in Figure 6a), while the peak current depends on the difference $\lambda_2 - \lambda_1$.

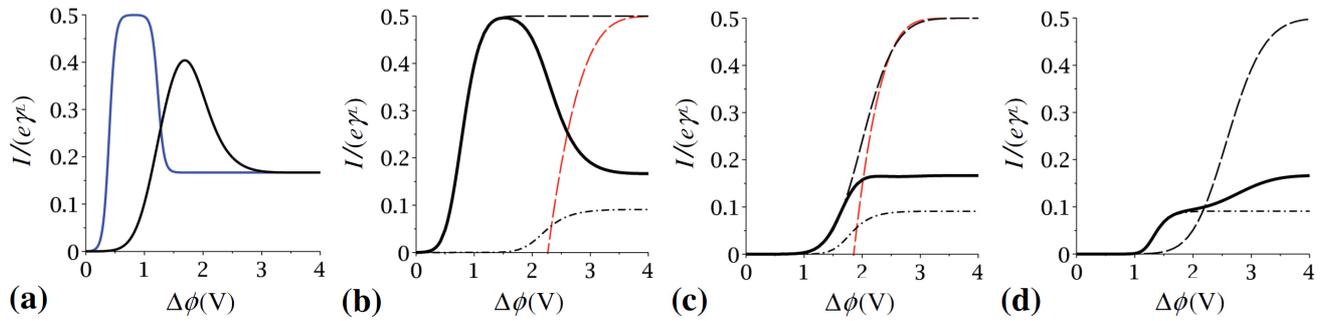

**Figure 6.** $I/(e\gamma^L)$ versus $\Delta\phi$, obtained by means of eq 14 (blue and black full lines), for a two-channel junction with different values of the reorganization energies associated with the two channels. The parameters are: $\varepsilon_1 - \mu = 0.2\,\text{eV}$, $\varepsilon_2 - \mu = 0.6\,\text{eV}$, $\gamma_1^L = \gamma_1^R = \gamma_2^L = 10\gamma_2^R$, $-\phi_L = \phi_R = \Delta\phi/2$, $T = 298\,\text{K}$ and **(a)** blue line: $\lambda_1 = \lambda_2 = 0\,\text{eV}$; black line: $\lambda_1 = 0.5\,\text{eV}$, $\lambda_2 = 0.6\,\text{eV}$; **(b)** $\lambda_1 = 0.25\,\text{eV}$, $\lambda_2 = 0.8\,\text{eV}$; **(c)** $\lambda_1 = 0.9\,\text{eV}$, $\lambda_2 = 0.55\,\text{eV}$; **(d)** $\lambda_1 = 1.2\,\text{eV}$, $\lambda_2 = 0.25\,\text{eV}$. Black dashed line: $I_{AB}$ obtained from eq 11a. Black dash-dotted line: $I_{AC}$ obtained from the analog of eq 11a for the *AC* channel. $\frac{1}{2}\log_{10}(R_{AC}^L/R_{CA}^R)$ is also shown (red dashed line) in the voltage range where $R_{AC}^L \geq R_{CA}^R$.

In Figures 6a-b the two channels are accessed sequentially as established by eq 16 for the given model parameters. As the bias increases, $I$ grows as $I_{AB}$ before channel 2 is accessed. In this bias range, $R_{AC}^L$ is negligible and indeed smaller than $R_{CA}^R$ (see red-dashed line in panel b), because of the respective EETV values, as given by eqs 13a-b. Therefore, the molecular system has a negligible probability to be trapped in state $|C\rangle$. Above $\Delta\phi \sim 2.5\,\text{eV}$ channel 2 becomes accessible, and $R_{AC}^L$ increases and becomes similar in magnitude to the rates into and out of state $|B\rangle$. Once $R_{AC}^L/R_{CA}^R$ approaches its plateau value of $\gamma_2^L/\gamma_2^R = 10$, the transferring charge can be temporarily trapped in state $|C\rangle$, leading to decreasing current through the junction – manifestation of NDR. Thereafter, the steady-state current takes its plateau value arising from the competition between the two fully operating transport channels, as described by eq 25.

NDR does not occur in Figure 6c, although eq 24 is satisfied, because in this case the two conduction channels are not accessed sequentially. In fact, $I_{AB}$, $R_{AC}^L/R_{CA}^R$ and so $I_{AC}$ rise essentially in the same voltage range, and, as a result, the current smoothly reaches its high-voltage value without going through a maximum.

NDR does not occur also in the case of Figure 6d, where the current reaches the plateau value of $I_{AC}$ and then increases to its high-voltage plateau value. Here, because of the large reorganization energy that characterizes the $AB$ channel, the order in which the two channels are accessed is inverted relative to that expected from the values of $\varepsilon_1$ and $\varepsilon_2$ themselves. eq 24, recast in the form

$$\frac{1}{\gamma_1^R} > \frac{1}{\gamma_2^L} + \frac{1}{\gamma_2^R}, \tag{26}$$

is in fact not satisfied, in agreement with the absence of NDR in this case. Notice that the occurrence of cases such as shown in Figure 6d, where the access order of the two transport channels is determined by the corresponding reorganization energies rather than the bare energies $\varepsilon_1$ and $\varepsilon_2$, generally requires a significant difference between $\lambda_1$ and $\lambda_2$. In contrast, cases as seen in Figures 6a-c can commonly take place for similar values of $\lambda_1$ and $\lambda_2$.

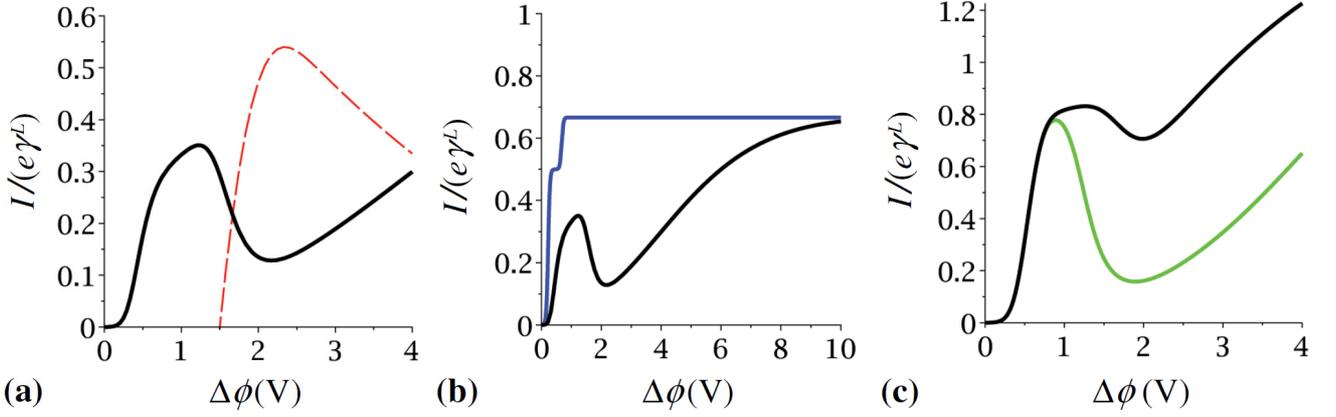

**Figure 7.** $I/(e\gamma^L)$ plotted against $\Delta\phi$ (eq 14) in the two-channel model with parameters: $T = 298\,\text{K}$, $-\phi_L = w\Delta\phi$ and $\phi_R = (1-w)\Delta\phi$ with $w = 0.9$, $\varepsilon_1 - \mu = 0.2\,\text{eV}$, $\varepsilon_2 - \mu = 0.6\,\text{eV}$, $\lambda_1 = 0.3\,\text{eV}$, and $\lambda_2 = 1.2\,\text{eV}$. In (a) and (b) the coupling strengths to the electrodes are given by $\gamma_1^L = \gamma_1^R = \gamma_2^L = \gamma_2^R$. The red dashed line in (a) displays $\tfrac{1}{2}\log_{10}(R_{AC}^L/R_{CA}^R)$. The black curve in panel b is reported from (a), while the blue curve corresponds to $\lambda_1 = \lambda_2 = 0$. In (c), $\gamma_1^R = \gamma_2^R = 9\gamma_1^L = 9\gamma_2^L$ and the characteristic in green color is obtained for $\varepsilon_2 - \mu = 0.4\,\text{eV}$.

Figure 7 shows scenarios unpredicted by either eq 24 or eq 26 that arise in the presence of $\lambda_1$ and $\lambda_2$. Here we assume an asymmetric potential distribution, taking the asymmetry parameter to be $w = 0.9$, usually implying that one lead is more strongly connected to the molecular bridge than its counterpart.[50] Since $w$ is close to unity and $\lambda_2$ is significantly larger than $\varepsilon_2 - \mu$, the second term in the brackets of eq 20a, rewritten for channel 2, implies a large threshold for the $C \to A$ transition at the $R$ contact. Thus, at low enough voltages, $I = I_{AB}$. In the bias voltage range $-\phi_L^{EETV(AC)}/w < \Delta\phi < \phi_R^{EETV(CA)}/(1-w)$, the transferring electron has an appreciable probability to be trapped in the molecular state $|C\rangle$, as quantified by the increase in the ratio $R_{AC}^L/R_{CA}^R$ (see panel a), and the current decreases, namely, NDR occurs. Note also that $R_{AC}^L/R_{CA}^R$ begins to be significant while $I_{AC}$ is still negligible. So, according to the discussion of eq 25, the current would be $I = I_{AC}$ for large enough $\lambda_2$.

For $\Delta\phi > \phi_R^{EETV(CA)}/(1-w)$, the forward $C \to A$ transition can also occur efficiently and the current can increase to its maximum high-voltage value that depends on the details of the molecular-leads coupling. For the coupling parameters used in the figure we have $I_{max} = \frac{4}{3}(I_{AB})_{max} = \frac{4}{3}(I_{AC})_{max}$ in panels a and b and $I_{max} = \frac{20}{11}(I_{AB})_{max} = \frac{20}{11}(I_{AC})_{max}$ in panel c, where the different coupling strengths do not lead to qualitatively new features. Current increase beyond the NDR as predicted in Figure 7 is indeed observed in many systems.[34–38, 58] In contrast, the mechanism implied by eq 24 or 26 predicts that the current plateaus at voltages above the NDR peak (see Appendix B).

The situations described in Figure 7 show that condition of eq 24 is not necessary for the occurrence of NDR in the presence of environmental relaxation. NDR occurs even though the *AB* and *AC* transport channels are accessed in distinct bias ranges (e.g., cf. the solid and dash-dot curves in Figure 7a). In agreement with the discussion of eq 20, the situations depicted in Figure 7 cannot occur for $\lambda_1 = \lambda_2 = 0$, which can give only the blue characteristic.

In such NDR scenarios the peak-to-valley current ratio increases with the separation of the bias ranges where the ET processes in the two channels are accessed. Moreover, the shape of the NDR phenomenon is particularly sensitive to changes in $\varepsilon_2 - \mu$. For example, reducing $\varepsilon_2 - \mu$ (see the green line in Figure 7) increases the EETV pertaining to $R_{CA}^R$ (which dictates the threshold voltage for access of the *AC* channel at high $w$ values), but also decreases the EETV pertaining to $R_{AC}^L$ and thus the bias at which NDR starts. This modulates the shape of the NDR region and can bring about a considerable increase in the peak-to-valley current ratio.

*4.2. Temperature-dependent NDR, controllable by the channel reorganization energies.* Other interesting NDR effects that may occur under asymmetric junction bias distribution are shown in Figure 8. Equal coupling strengths to each lead in the two conduction channels are assumed here, and hence NDR would not be predicted by eq 24. Still, it is observed in Figures 8a-b (another example is provided in Figure S2 of the Supporting Information). Here we discussed the temperature dependence of these observations. The following points are noteworthy:

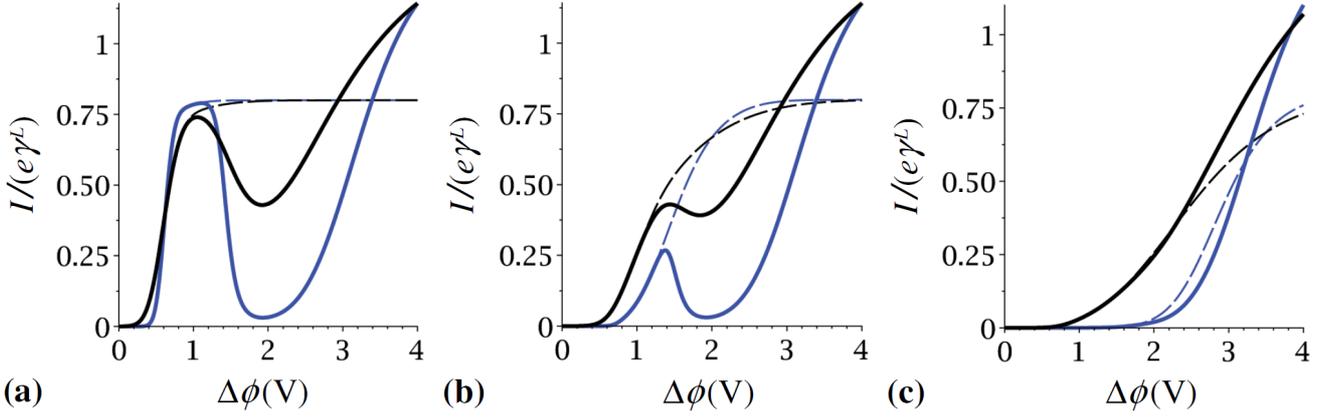

**Figure 8.** $I/(e\gamma^L)$ versus $\Delta\phi$ (eq 14) in the two-channel model system. The reorganization energy of the *AC* transport channel is fixed at the value $\lambda_2 = 1.2\,\text{eV}$, while that of the *AB* channel is given by **(a)** $\lambda_1 = 0.3\,\text{eV}$, **(b)** $\lambda_1 = 0.6\,\text{eV}$, and **(c)** $\lambda_1 = 0.9\,\text{eV}$. The $I$-$\Delta\phi$ characteristics in black and blue correspond to $T = 298\,\text{K} \equiv T_0$ and $T_0/3$, respectively. The other parameters are: $-\phi_L = w\Delta\phi$ and $\phi_R = (1-w)\Delta\phi$ with $w = 0.8$, $\gamma_1^R = \gamma_2^R = 4\gamma_1^L = 4\gamma_2^L$, $\varepsilon_1 - \mu = 0.2\,\text{eV}$ and $\varepsilon_2 - \mu = 0.4\,\text{eV}$. Black dashed line: $I_{AB}$.

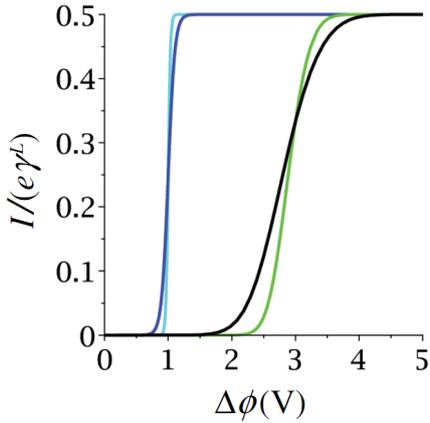

**Figure. 9.** $I/(e\gamma^L)$ versus $\Delta\phi$, (eq 14) for $-\phi_L = \phi_R = \Delta\phi/2$, $\gamma_1^L = \gamma_1^R$, and $\varepsilon_1 - \mu = 0.5\,\text{eV}$. The black ($\lambda_1 = 1\,\text{eV}$) and blue ($\lambda_1 = 0\,\text{eV}$) lines correspond to $T = T_0 = 298\,\text{K}$ while the green and cyan ones are respectively obtained for $T = T_0/3$.

(a) The occurrence of this NDR does not depend critically on the values of the couplings to the leads;

(b) The NDR peak height can either decrease (as in panel a) or increase (panel b) with temperature, depending on the proximity of $\lambda_1$ to $\lambda_2$.

(c) In both cases, the peak to valley current ratio decreases with increasing temperature.

(d) We have noted above that the nuclear reorganization smoothes the voltage dependence of the current. As is clear from eq 7 and expected on physical grounds, this is also caused by a temperature increase.

NDR behaviors as in Figures 8a and 8b have been both observed in different experiments.[35–37] In order to understand these two different temperature dependencies of the NDR peak, it is useful to consider the simple case of transport through a single channel with $\lambda_1 = 0$, illustrated in Figure 9. For definiteness, we assume that at zero bias the molecular level lies above the electrodes' Fermi level: $\varepsilon_1 - \mu > 0$. Under the kinetic scheme considered here, at low enough voltages an increase in $T$ will enhance electron transport through this level. On the contrary, at high biases, when the molecular level is within the Fermi window, the current decreases with increasing $T$, as it can be easily realized by considering the Fermi populations involved.

This observation changes quantitatively, but not qualitatively, when the reorganization energy is finite, as it is displayed in Figure 9. This is also the temperature dependence of $I \cong I_{AB}$ before the onset of NDR in Figures 8-b. Depending on the relative values of $\lambda_1$ and $\lambda_2$, hence on the differences between the EETV values corresponding to the two channels, the departure of $I$ from $I_{AB}$ and its consequent peak can occur over a voltage range where $I_{AB}$ is affected either positively or negatively by temperature increase. This explains the opposite trends for the intensity of the current peak in Figures 8a and 8b. On the other hand, in both cases the NDR and the consequent minimum in the current occur in a bias voltage range where $I_{AC}$ increases with temperature, which entails less charge trapping in state $|C\rangle$ and a smaller peak-to-valley current ratio with increasing temperature.

No NDR is obtained with the choice of $\lambda_1$ and $\lambda_2$ in Figure 8c. Altogether, Figures. 8a-c suggest a strategy for controlling the occurrence, realization (in particular, the voltage and the intensity of the peak current) and temperature dependence (peak shift and suppression or enhancement) of NDR, based on the relative values of $\lambda_1$ and $\lambda_2$, that is, on the composition and operation of the (solvated) bridge system. The proposed mechanism does not make any specific assumptions on the nature of the molecular bridge and the electrodes, except for a few parameters that globally characterize the energetics of the molecular system ($\varepsilon_1$, $\varepsilon_2$, $\lambda_1$, and $\lambda_2$). Such a generality may have important implications for practical purposes.

*4.3. Effect of backward electron transitions.* In this section we examine NDR effects that are strictly related to the voltage-dependent backward electron transitions starting from a blocking state (i.e., a molecular state that can be filled by ET from the *L* contact but cannot be emptied via ET to the *R* contact because its coupling to this contact is negligible).

Assume that $|C\rangle$ is the blocking state. This means that $\gamma_2^R \cong 0$, and thus $R_{CA}^R$ and $R_{AC}^R$ are negligible throughout the explored voltage range. On the other hand, the electron transitions between the molecule in state $|C\rangle$ and the *L* metal can have appreciable rates $R_{AC}^L$ and $R_{CA}^L$ over a suitable bias voltage range that depends on the nature of the system. Moreover, according to the analog of eq 9 for channel 2 and the *L* contact, it is

$$R_{AC}^L / R_{CA}^L = \exp\left(\frac{\mu - \varepsilon_2 - e\phi_L}{k_B T}\right), \tag{27}$$

so that the $C \to A$ transition is faster than the reverse transition for $\Delta\phi = -\phi_L/w < (\varepsilon_2 - \mu)/(ew)$. If the current through channel 1 becomes appreciable in this voltage range, $|C\rangle$ does not behave as a trapping state, even if $R_{AC}^L$ and $R_{CA}^L$ are both small.[59] Instead, for $\Delta\phi > (\varepsilon_2 - \mu)/(ew)$, the $C \to A$ transition at the *L* interface is forbidden (i.e., $|C\rangle$ starts to act as a trapping state) and NDR takes place. This NDR is not predicted by eq 15, which neglects the backward ET processes in both channels. On the other hand, as seen in section 2.1, the backward ET rates $R_{BA}^L$ and $R_{AB}^R$ are negligible at bias voltages for which $I_{AB}$ is appreciable. Ultimately, the NDR mechanism under consideration can be described by retaining only the rates $R_{AB}^L$, $R_{BA}^R$, $R_{AC}^L$, and $R_{CA}^L$ in eq 14, which leads to

$$I \cong \frac{I_{AB}}{1 + \dfrac{R_{BA}^R}{R_{AB}^L + R_{BA}^R} \exp\left(\dfrac{\mu - \varepsilon_2 - e\phi_L}{k_B T}\right)}, \tag{28}$$

where $I_{AB}$ is given by eq 12. Results based on eq 28 are displayed in Figure 10. In particular, Figure 10a shows a successful application of this equation in a common case where eq 15 fails to describe the correct current-voltage characteristic. Thus, eq 28 can be a useful and simple analytical expression for the fitting and interpretation of some phenomena of current collapse and NDR that are missed if effects of backward transitions are disregarded altogether.

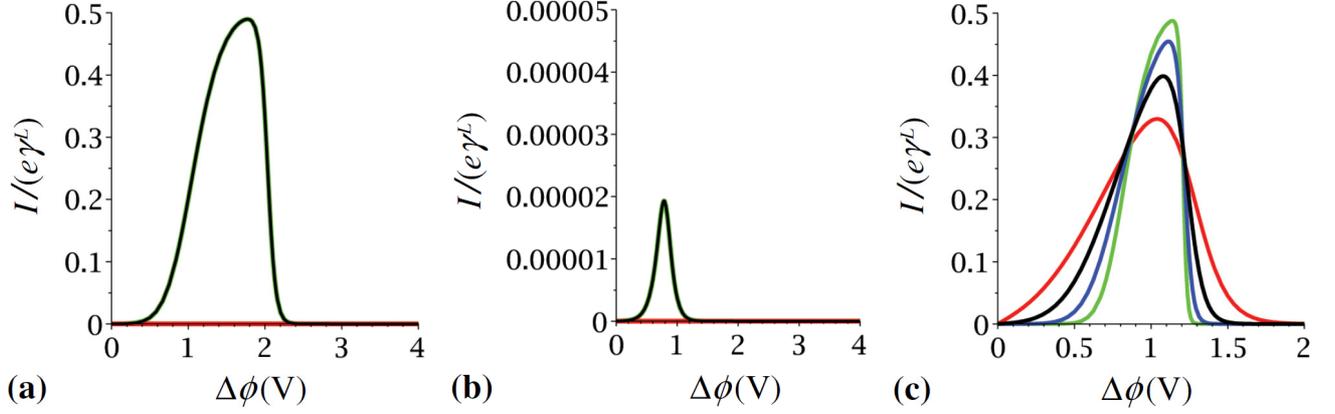

**Figure 10.** $I/(e\gamma^L)$ versus $\Delta\phi$, from eq 14 (solid black) and the corresponding approximations, eq 15 (red) and eq 28 (green), except for panel c, where only eq 14 is used. Parameters are: $T = T_0 = 298\,\text{K}$, $-\phi_L = \phi_R = \Delta\phi/2$, $\gamma_1^L = \gamma_1^R = \gamma_2^L$, $\gamma_2^R = 0$, and **(a)** $\varepsilon_1 - \mu = 0.2\,\text{eV}$, $\varepsilon_2 - \mu = 1.0\,\text{eV}$, $\lambda_1 = \lambda_2 = 0.4\,\text{eV}$; **(b)** $\varepsilon_1 - \mu = 0.2\,\text{eV}$, $\varepsilon_2 - \mu = 0.4\,\text{eV}$, $\lambda_1 = 1.2\,\text{eV}$, $\lambda_2 = 1.4\,\text{eV}$; **(c)** $\varepsilon_1 - \mu = 0.1\,\text{eV}$, $\varepsilon_2 - \mu = 0.6\,\text{eV}$, $\lambda_1 = 0.35\,\text{eV}$, $\lambda_2 = 0.7\,\text{eV}$, and temperatures $T_0$ (black), $2T_0$ (red), $T_0/2$ (blue), and $T_0/4$ (green).

The above analysis indicates that NDR occurs if the threshold bias voltage for appreciable current via channel 1 is smaller than $(\varepsilon_2 - \mu)/(ew)$. In fact, NDR is obtained in Figure 10a, while it is virtually absent in Figure 10b, where this condition is not satisfied. Notice also that according to eq 28 the current is cut off within a voltage range of a few thermal voltages $k_B T/e$ (as it is observed in many experiments[33, 35]), once $\Delta\phi > (\varepsilon_2 - \mu)/(ew)$. Thus, the lower is $T$, the sharper is the descent of the $I$-$\Delta\phi$ characteristic (see Figure 10c), namely, the more pronounced is the NDR peak.

It is also interesting to compare the NDR mechanisms operating in cases like those of Figure 6b and Figure 10a. In the first case, the crossing between the values of $R_{AC}^L$ and $R_{CA}^L$ occurs at a voltage where $R_{CA}^R$ dominates the $C \to A$ transition rate, as $R_{CA}^R$ increases with the voltage and $\gamma_2^R$ is not negligible (even though $\gamma_2^L > \gamma_2^R$). Consequently, the current-voltage response is dominated by the forward ET rates $R_{AC}^L$ and $R_{CA}^R$, the NDR region begins at biases such that $R_{AC}^L > R_{CA}^R$, and the decrease in the current is much less sharp than that in Figures 10a-b. On the contrary, when $|C\rangle$ is a blocking state, the role played by the backward transition of rate $R_{CA}^L$ on the $I$-$\Delta\phi$ response is accentuated, and is visible in the shape of the NDR peak.

The role of the C → A backward transitions in causing NDR, discussed in this subsection, cannot be grasped by the analysis[30] leading to eq 24, because this equation is obtained from the plateau currents for the one-channel and two-channel transport regimes over voltage ranges where all backward electron transitions are disregarded (see also Appendix B).

## 5. Conclusions.

This article has focused on non-linear transport properties of molecular conduction junctions as revealed from a Marcus-level theory applied to a minimal model of a molecular junction characterized by the following attributes: (a) Electronic conduction is dominated by two charging states of the bridge molecule, referred to as a neutral and a negative ion species, weakly coupled to metal electrodes. (b) Two states (e.g., the ground and first-excited states of the isolated molecule) of the charged species are involved in conduction in the voltage range of interest, implying two conduction channels. (c) These states are characterized by different electron localization properties, an effect that is further enhanced by environmental relaxation. (d) The coupling between molecule and metal leads is weak enough to allow analysis based on Marcus theory. The latter attribute is technical, and can be relaxed in more advanced treatments.

The resulting generic model shows pronounced non-linear transport behavior and provide an alternative rationalization of many observations that are otherwise interpreted as voltage induced conformational changes. Both switching mechanisms are expected to exist in different molecular conduction situations, and they can be distinguished by their response to gating.

In the Marcus-level approach, valid in the weak molecule-lead couplings' limit and at relatively high temperatures, the current is obtained in terms of the rate constants for the individual molecule-electrode ET processes. These rates can be expressed in simple analytical forms that allows for exploration of different approximations and limits. The resulting kinetic framework leads to a rich phenomenology of non-linear transport phenomena. In particular, the following aspects of transport in this regime were discussed:

(1) The dependence of the current-voltage response on molecular and, especially, solvent relaxation, as expressed by the reorganization energies associated with transitions between the different molecular electronic states, was elucidated

(2) The important effect of nuclear reorganization on current rectification phenomena associated with asymmetric voltage distribution across the junction was pointed out.

(3) The role played by nuclear reorganization in determining the existence and the onset of negative differential conductance through such junctions was investigated and clarified. The main effect of nuclear relaxation was shown to be the redefinition of accessibility criteria to the different conduction channels involved in the three-state model.

(4) The current-voltage behavior beyond the NDR regime was investigated. In particular, current increase at higher voltages, as shown in Figures 7-8, is strictly related to the interplay between the potential distribution across the junction and the presence of nuclear relaxation.

(5) When NDR is seen, its dependence on temperature, has been studied. Room temperature or even higher temperature NDR are found, in agreement with many recent observations in diverse nanodevices.[34–39] Our theoretical analysis shows that the NDR peak in the current-voltage characteristic can either increase or decrease with temperature, whereas the peak/valley ratio always decreases with increasing temperature. Such conclusions are in general agreement with a variety of experimental results reported in the literature (e.g., see refs 33, 35, and 37).

(6) Approximate expressions previously employed to describe only plateau currents[30] were shown to describe well the full current-voltage behavior in many situations. Cases where such expressions fail were also highlighted and clarified.

The Marcus theory based kinetic framework used in this work is valid under well known conditions, in particular, the weak molecule-electrode coupling condition that implies that local thermal equilibrium is achieved on a timescale fast relative to electron-transfer rates. It should be pointed out, however, that this condition is not always simple in present contexts, because the coupling itself changes with the evolution of solvent response. A fully equilibrated molecular electronic state can be weakly coupled to a nearby metal even if the bare electronic coupling is large, because of the involvement of nuclear matrix elements (Franck Condon factors) that evolve during the solvation/relaxation process. This makes the analysis of intermediate cases particularly hard in such systems. Here we avoided this difficult regime and focused on situations that are characterized by weak coupling kinetics.

The generic nature of our model may limit its applicability to analyzing detailed properties of individual systems. On the other hand, the generic origin of the properties investigated, in particular the important phenomenon of negative differential resistance, and its dependence on just a few simple parameters is potentially important for nanotechnology applications. It is worth noting that our discussion in sections 3-4 offers some hints on the physical parameters that can be tuned in order to get different behaviors regarding NDR; e.g., see the effect of the $\lambda_1$ value, which depends on the choice of the bridge, in Figure 8.

Because of its simplicity and generic nature, the kinetic description used in the present paper provides a useful framework for further investigations on electrochemical redox reactions at metal electrodes and electrochemical redox junctions. In a subsequent paper we study multistability and hysteresis phenomena as described by within the same framework.

**Acknowledgement.** The research of A.N. is supported by the Israel Science Foundation, the Israel-US Binational Science Fundation, the European Science Council (FP7 /ERC grant no. 226628) and the Israel – Niedersachsen Research Fund. We wish to thank Spiros S. Skourtis and Philip Schiff for helpful discussions.

**Appendix A. Analytical derivation of eqs 7a-d.**

In this appendix we present the mathematical derivation of eqs 7a-d. First, we wish to point out the reasons for our choice to characterize the potential distribution across the molecular bridge by means of an effective potential. In redox molecules, the localization of excess charge on a redox center allows to conveniently describe the ET to and from the metal contacts by assigning a given effective (e.g. average) electrostatic potential to the relatively small region (as compared to the size of the molecule) where the excess electron is localized. For example, this consideration applies to the molecular systems mentioned in section 3, where the side chain of a monomer is replaced by a redox group. In non-redox molecules with large capacitance,[60] the transferring excess electron is expected to spread over the whole molecule, thus yielding a rather uniform contribution to the electrostatic potential distribution, as recently shown via a density-functional theory approach.[60, 61] Also such molecules can be well described by means of a single potential value. In this work, we use a model where an effective potential describes the molecular bridge in a given charge transport channel, by considering it as a "zero-order" approximation to any case that can arise in molecular junctions.

Let us now consider the expression of $R_{AB}^K$ in eq 6a, where $\phi_K$ ($K = L$ or $R$) is the potential of the electrode relative to the molecular system. We separate the integrations over the positive and negative $E$ ranges, change the integration variable in the integral over the negative energy range to $-E$ and rename it $E$, rearrange the exponentials in the Fermi functions, and exploit the summation rule of the geometric series, thus obtaining the following:

$$R_{AB}^K = \frac{\gamma^K}{2\sqrt{\pi \lambda k_B T}} \left\{ \int_0^\infty dE \frac{\exp\left(-\dfrac{E}{k_B T}\right)}{1+\exp\left(-\dfrac{E}{k_B T}\right)} \exp\left[-\frac{(E+\alpha_K-\lambda)^2}{4\lambda k_B T}\right] \right.$$

$$\left. + \int_0^\infty dE \frac{1}{1+\exp\left(-\dfrac{E}{k_B T}\right)} \exp\left[-\frac{(-E+\alpha_K-\lambda)^2}{4\lambda k_B T}\right] \right\}$$

$$= \frac{\gamma^K}{2\sqrt{\pi \lambda k_B T}} \exp\left[-\frac{(\alpha_K-\lambda)^2}{4\lambda k_B T}\right] \left\{ \int_0^\infty dE \frac{\exp\left(-\dfrac{E}{k_B T}\right)}{2-\left[1-\exp\left(-\dfrac{E}{k_B T}\right)\right]} \exp\left(-\frac{E^2}{4\lambda k_B T} - \frac{\alpha_K-\lambda}{2\lambda k_B T}E\right) \right.$$

$$+ \int_0^\infty dE \frac{1}{2 - \left[1 - \exp\left(-\frac{E}{k_B T}\right)\right]} \exp\left(-\frac{E^2}{4\lambda k_B T} + \frac{\alpha_K - \lambda}{2\lambda k_B T} E\right) \Bigg\}$$

$$= \frac{\gamma^K}{4\sqrt{\pi \lambda k_B T}} \exp\left[-\frac{(\alpha_K - \lambda)^2}{4\lambda k_B T}\right] \Bigg\{ \int_0^\infty dE \frac{1}{1 - \left[1 - \exp\left(-\frac{E}{k_B T}\right)\right]/2} \exp\left(-\frac{E^2}{4\lambda k_B T} - \frac{\lambda + \alpha_K}{2\lambda k_B T} E\right)$$

$$+ \int_0^\infty dE \frac{1}{1 - \left[1 - \exp\left(-\frac{E}{k_B T}\right)\right]/2} \exp\left(-\frac{E^2}{4\lambda k_B T} - \frac{\lambda - \alpha_K}{2\lambda k_B T} E\right) \Bigg\} \quad (A1)$$

$$= \frac{\gamma^K}{4\sqrt{\pi \lambda k_B T}} \exp\left[-\frac{(\alpha_K - \lambda)^2}{4\lambda k_B T}\right] \int_0^\infty dE \sum_{n=0}^\infty \frac{1}{2^n} \left[1 - \exp\left(-\frac{E}{k_B T}\right)\right]^n$$

$$\times \left[\exp\left(-\frac{E^2}{4\lambda k_B T} - \frac{\lambda + \alpha_K}{2\lambda k_B T} E\right) + \exp\left(-\frac{E^2}{4\lambda k_B T} - \frac{\lambda - \alpha_K}{2\lambda k_B T} E\right)\right],$$

where the notations $\lambda \equiv \lambda_1$ and $\gamma^K \equiv \gamma_1^K$ are used for the sake of simplicity. Further elaboration requires the interchange of the integration and summation operations. This is appropriately allowed by Lebesgue integration theory even in the presence of improper integrals.[62] However, the extension of the integration interval to $+\infty$ is, indeed, an approximation.[41] Moreover, the summation needs to be truncated at a finite value of $n$ in any application of eq A1, because the analytic form of the sum of the series is not known. Therefore, we can anyway truncate the series to a value of $N$ that is large enough for the convergence of the results in the given system and write

$$R_{AB}^K = \frac{\gamma^K}{4\sqrt{\pi \lambda k_B T}} \exp\left[-\frac{(\alpha_K - \lambda)^2}{4\lambda k_B T}\right] \sum_{n=0}^N \frac{1}{2^n} \sum_{j=0}^n (-1)^j \binom{n}{j}$$
$$\times \int_0^\infty dE \left\{\exp\left[-\frac{E^2}{4\lambda k_B T} - \frac{(2j+1)\lambda + \alpha_K}{2\lambda k_B T} E\right] + \exp\left[-\frac{E^2}{4\lambda k_B T} - \frac{(2j+1)\lambda - \alpha_K}{2\lambda k_B T} E\right]\right\}. \quad (A2)$$

The integral in eq A2 is given by the formula[63]

$$\int_0^\infty \exp\left(-\frac{x^2}{4\beta} - \gamma x\right) dx = \sqrt{\pi \beta} \exp(\beta \gamma^2)[1 - \text{erf}(\gamma \sqrt{\beta})] = \sqrt{\pi \beta} \exp(\beta \gamma^2) \text{erfc}(\gamma \sqrt{\beta}), \quad (\text{Re}\,\beta > 0) \quad (A3)$$

so that the first equation 7a is finally obtained. Analogously, for $R_{BA}^K$ we obtain

$$\begin{aligned}
R_{BA}^K &= \frac{\gamma^K}{2\sqrt{\pi \lambda k_B T}} \left\{ \int_0^\infty dE \frac{1}{1+\exp\left(-\frac{E}{k_B T}\right)} \exp\left[-\frac{(E+\alpha_K+\lambda)^2}{4\lambda k_B T}\right] \right. \\
&\quad \left. + \int_0^\infty dE \frac{1}{1+\exp\left(\frac{E}{k_B T}\right)} \exp\left[-\frac{(-E+\alpha_K+\lambda)^2}{4\lambda k_B T}\right] \right\} \\
&= \frac{\gamma^K}{2\sqrt{\pi \lambda k_B T}} \left\{ \int_0^\infty dE \frac{1}{1+\exp\left(-\frac{E}{k_B T}\right)} \exp\left[-\frac{(E+\alpha_K+\lambda)^2}{4\lambda k_B T}\right] \right. \\
&\quad \left. + \int_0^\infty dE \frac{\exp\left(-\frac{E}{k_B T}\right)}{\exp\left(-\frac{E}{k_B T}\right)+1} \exp\left[-\frac{(-E+\alpha_K+\lambda)^2}{4\lambda k_B T}\right] \right\} \\
&= \frac{\gamma^K}{4\sqrt{\pi \lambda k_B T}} \exp\left[-\frac{(\alpha_K+\lambda)^2}{4\lambda k_B T}\right] \int_0^\infty dE \frac{1}{1-\left[1-\exp\left(-\frac{E}{k_B T}\right)\right]/2} \\
&\quad \times \left[ \exp\left(-\frac{E^2}{4\lambda k_B T} - \frac{\lambda+\alpha_K}{2\lambda k_B T}E\right) + \exp\left(-\frac{E^2}{4\lambda k_B T} - \frac{\lambda-\alpha_K}{2\lambda k_B T}E\right) \right] \\
&= \frac{\gamma^K}{4} \exp\left[-\frac{(\alpha_K+\lambda)^2}{4\lambda k_B T}\right] \sum_{n=0}^N \frac{1}{2^n} \sum_{j=0}^n (-1)^j \binom{n}{j} \left[\chi_j(\lambda,T,\alpha_K) + \chi_j(\lambda,T,-\alpha_K)\right] \quad (A4) \\
&= R_{AB}^K \frac{\exp\left[-\frac{(\alpha_K+\lambda)^2}{4\lambda k_B T}\right]}{\exp\left[-\frac{(\alpha_K-\lambda)^2}{4\lambda k_B T}\right]} = R_{AB}^K \exp\left(-\frac{\alpha_K}{k_B T}\right) = R_{AB}^K \exp\left(-\frac{\mu_K - \varepsilon_1 - e\phi_K}{k_B T}\right),
\end{aligned}$$

that is the second equation 7a. Hale's approximation[45] amounts to include only $\chi_0(\lambda,T,-\alpha_L)$ in $R_{AB}^L$ and $\chi_0(\lambda,T,\alpha_R)$ in $R_{BA}^R$, which leads to

$$R_{AB}^L \cong \frac{\gamma^L}{2} \operatorname{erfc}\left(\frac{\lambda-\alpha_L}{2\sqrt{\lambda k_B T}}\right) \quad (A5)$$

and

$$R_{BA}^{R} \cong \frac{\gamma^{R}}{2}\operatorname{erfc}\left(\frac{\lambda + \alpha_{R}}{2\sqrt{\lambda k_{B}T}}\right), \quad (A6)$$

respectively. Indeed, the detailed analysis of eqs 7a-d which is reported in the Supporting Information identifies the limits of applicability of eqs A5 and A6, therefore clarifying and justifying their use under weaker physical conditions than the ones implicit in Hale's derivation. In short, eqs A5-6 can be used at high enough bias voltages, or at any voltage if the reorganization energy is sufficiently large. Note that eqs 8a and 8b result immediately from eqs A5 and A6, respectively.

**Appendix B. Current expressions in the two-state and three-state models.**

When the molecular bridge is modeled as a two-state system, the steady-state expression of the master equation is given by eq 10 and brings about the bias-dependent state probabilities

$$\begin{cases} P_A = \dfrac{R_{BA}}{R_{AB} + R_{BA}} \\ P_B = \dfrac{R_{AB}}{R_{AB} + R_{BA}}. \end{cases} \quad (B1)$$

The steady-state current $I$, which equals both the left and right terminal currents, is given by

$$\frac{I}{e} = P_A R_{AB}^{L} - P_B R_{BA}^{L} = -P_A R_{AB}^{R} + P_B R_{BA}^{R} = \frac{R_{AB}^{L} R_{BA}^{R} - R_{BA}^{L} R_{AB}^{R}}{R_{AB} + R_{BA}}. \quad (B2)$$

Both the terms in the numerator of eq B2 must be considered at small biases, i.e., when $\Delta\phi$ is smaller than or comparable with $V_T$. For example, at zero bias the current is zero since eq 9 gives

$$R_{BA}^{L}(0)R_{AB}^{R}(0) = R_{AB}^{L}(0)\exp\left(-\frac{\alpha}{k_{B}T}\right)R_{BA}^{R}(0)\exp\left(\frac{\alpha}{k_{B}T}\right) = R_{AB}^{L}(0)R_{BA}^{R}(0), \quad (B3)$$

where $\alpha \equiv \mu - \varepsilon_1$ and the dependence of the rates on the potential drop across the junction has been shown explicitly. Moreover, the initial linear dependence of the current on the voltage is also obtained from the complete expression for the current. In fact, up to terms linear in $\Delta\phi$, it is

$$\frac{I(\Delta\phi)}{e} \cong \frac{R_{AB}^L(0)R_{BA}^R(0)}{R_{AB}(0)+R_{BA}(0)} \frac{\Delta\phi}{V_T}. \tag{B4}$$

Let us now consider the case where three molecular electronic states are involved in the transport. The steady state master equation is

$$\begin{cases} \dfrac{dP_A}{dt} = -P_A(R_{AB}+R_{AC}) + P_B R_{BA} + P_C R_{CA} = 0 \\ \dfrac{dP_B}{dt} = P_A R_{AB} - P_B R_{BA} = 0 \\ \dfrac{dP_C}{dt} = P_A R_{AC} - P_C R_{CA} = 0. \end{cases} \tag{B5}$$

The resulting bias-dependent probabilities, subject to the normalization condition $P_A + P_B + P_C = 1$, are

$$P_A = \frac{1}{1+\dfrac{R_{AB}}{R_{BA}}+\dfrac{R_{AC}}{R_{CA}}}; \quad P_B = \frac{R_{AB}}{R_{BA}} P_A; \quad P_C = \frac{R_{AC}}{R_{CA}} P_A. \tag{B6}$$

The steady-state current $I$ equals the currents at both the left and right contacts. Using the expression for the left terminal current, we obtain

$$\begin{aligned}
\frac{I}{e} &= P_A R_{AB}^L - P_B R_{BA}^L + P_A R_{AC}^L - P_C R_{CA}^L = P_A\left(R_{AB}^L + R_{AC}^L - \frac{R_{AB}}{R_{BA}} R_{BA}^L - \frac{R_{AC}}{R_{CA}} R_{CA}^L\right) \\
&= \frac{1}{1+\dfrac{R_{AB}}{R_{BA}}+\dfrac{R_{AC}}{R_{CA}}}\left(\frac{R_{AB}^L R_{BA}^R - R_{AB}^R R_{BA}^L}{R_{AB}+R_{BA}}\frac{R_{AB}+R_{BA}}{R_{BA}} + \frac{R_{AC}^L R_{CA}^R - R_{AC}^R R_{CA}^L}{R_{AC}+R_{CA}}\frac{R_{AC}+R_{CA}}{R_{CA}}\right) \\
&= \frac{1}{e}\frac{1}{1+\dfrac{R_{AB}}{R_{BA}}+\dfrac{R_{AC}}{R_{CA}}}\left[I_{AB}\left(1+\frac{R_{AB}}{R_{BA}}\right) + I_{AC}\left(1+\frac{R_{AC}}{R_{CA}}\right)\right],
\end{aligned} \tag{B7}$$

where $I_{AB}$ and $I_{AC}$ are given by eq B2 as applied to the *AB* and *AC* transport channels, respectively. Finally, insertion of eq 9 and the corresponding equations for the *AC* conduction mode into eq B7 leads to eq 14 in the main text.

In ref 30, sequential access of the two transport channels is assumed. Then, the master equation B5 is solved in a bias range where channel 2 has not yet been accessed, which leads to the state probabilities in eq B1. At positive voltages such that the steady-state plateau current via channel 1, $I_{p1}$, is attained, the backward ET rates can be disregarded and the forward ET rates are given by the pertinent molecule-metal coupling strengths, so that eq 12 is obtained and the plateau current is written as

$$\frac{I_{p1}}{e} = \frac{\gamma_1^L \gamma_1^R}{\gamma_1^L + \gamma_1^R} . \tag{B8}$$

The solution of the full system (B5) under the same assumptions as above gives the final plateau current in the form of eq 15, hence

$$\frac{I_{p2}}{e} = \frac{\gamma_1^L + \gamma_2^L}{1 + \frac{\gamma_1^L}{\gamma_1^R} + \frac{\gamma_2^L}{\gamma_2^R}} . \tag{B9}$$

NDR occurs if $I_{p1} > I_{p2}$, which, by insertion of eqs B8 and B9, yields the condition of eq 24.[30]